# Design Constraints of Disturbance Observer-based Motion Control Systems are Stricter in the Discrete-Time Domain

Emre Sariyildiz
School of Mechanical, Materials, Mechatronic and Biomedical Engineering
Faculty of Engineering and Information Sciences, University of Wollongong,
Northfields Avenue Wollongong NSW 2522 Australia
emre@uow.edu.au

***Abstract*—This paper shows that the design constraints of the Disturbance Observer (DOb) based robust motion control systems become stricter when they are implemented using computers or microcontrollers. The stricter design constraints put new upper bounds on the plant-model mismatch and the bandwidth of the DOb, thus limiting the achievable robustness against disturbances and the phase-lead effect in the inner-loop. Violating the design constraints may yield severe stability and performance issues in practice; therefore, they should be considered in tuning the control parameters of the robust motion controller. This paper also shows that continuous-time analysis methods fall-short in deriving the fundamental design constraints on the nominal plant model and the bandwidth of the digital DOb. Therefore, we may observe unexpected stability and performance issues when tuning the control parameters of the digital robust motion controllers in the continuous-time domain. To improve the robust stability and performance of the motion controllers, this paper explains the fundamental design constraints of the DOb by employing the generalised continuous and discrete Bode Integral Theorems in a unified framework. Simulation and experimental results are given to verify the proposed analysis method.**

## I. Introduction

The DOb has been one of the most widely used motion control tools since it was proposed by K. Ohnishi in 1983 [1, 2]. In the last three decades, engineers and researchers have employed the DOb in many different motion control applications [2]. For example, it has been applied to i) precise positioning problems to improve the robustness of servo systems against disturbances [3 – 5], ii) physical robot-environment interaction tasks to estimate contact forces via Reaction Torque Observer (RTOb) without using a force/torque sensor [6 – 8], and iii) teleoperated robots to detect time-delay between master and slave robotic systems [9, 10]. The examples of high-performance motion control applications have motivated many control theoreticians and practitioners to build advanced analysis and synthesis tools for the DOb and expand its application areas [2, 11 - 15].

Although the DOb-based robust motion controller was originally proposed using Gopinath's observer synthesis method (aka. auxiliary variable-based observer design method) in state space [1, 2, 16], it started to receive increasing attention with the established frequency domain analysis and synthesis techniques, particularly from control engineering practitioners [2]. Over the past decades, various classical frequency-based analysis and synthesis methods have been proposed for the DOb-based motion control systems, e.g., loop shaping control methods such as Bode/Nyquist plots and $H\infty$ control and parametric uncertainty-based control [16 – 20].

Although the robust motion controllers are always implemented using computers or micro-controllers, analysis and synthesis are generally conducted in the continuous-time domain due to simplicity [2]. However, continuous-time analysis methods fall-short in deriving the fundamental design constraints on the design parameters of the DOb-based robust motion control systems implemented by digital controllers. This may cause some unexpected stability and performance problems in practice [21 – 23]. For example, the digital robust position controller exhibits oscillatory and unstable responses as the bandwidth of the DOb increases, and this dynamic behaviour cannot be explained in the continuous-time domain [21]. Despite some attempts on explaining the stability of the DOb-based digital robust motion controllers, the fundamental design constraints on the nominal plant model and the bandwidth of the DOb have yet to be clearly discussed in the literature [24, 25].

To conduct high-performance motion control applications via DOb, the control parameters (viz. the bandwidth of the observer, nominal plant parameters, and the outer-loop performance controller) of the robust motion control systems should be properly tuned [23, 26, 27]. When the design constraints of the DOb are violated, the robust stability and performance of the motion controller may notably deteriorate [27]. Therefore, it is essential to determine the upper and lower bounds on the bandwidth of the DOb, nominal plant parameters, and outer-loop performance controller. It is a well-known fact that the bandwidth of the DOb (i.e., the robustness against disturbances) is limited by noise sensitivity [20]. As the bandwidth of the DOb is increased, the robustness against disturbances improves yet the motion controller becomes more sensitive to the noise of velocity measurement. This, however, is not the only design constraint on the bandwidth of the DOb. The achievable bandwidth is also limited by the fundamental constraint of feedback control systems. Moreover, this also puts an upper bound on the plant model mismatch, which limits the achievable phase-lead effect in the inner-loop [27].

This paper proposes a new analysis method to explain the design constraints of the DOb-based motion control systems in the continuous- and discrete- time domains. By employing the generalised Bode Integral Theorems in a unified framework [28], it shows that the design constraints on the nominal plant model and the bandwidth of the DOb become stricter when the robust motion controller is implemented using computers and/or microcontrollers. The stricter design constraints cannot be explained using continuous-time analysis methods, which may lead to severe stability and performance problems in real motion control applications. Therefore, the control parameters of the DOb-based robust motion controllers should be tuned using discrete-time analysis and synthesis methods.

The rest of the paper is organised as follows. In section II, the DOb-based robust motion control is briefly introduced. In section III, the design constraints of the DOb are analysed in the continuous- and discrete- time domains. In section IV, simulation and experimental results are presented. In section V, the paper ends with conclusion.

## II. Disturbance Observer based Robust Motion Control

This section briefly introduces the DOb-based control and its robust position and force control applications in the field of motion control.

### A. Disturbance Observer

Figure 1 illustrates the block diagram of the DOb in the continues-time and discrete-time domains [16, 27]. In this figure, $J_m$ and $J_{m_n}$ are the uncertain and nominal motor inertias, respectively; $K_\tau$ and $K_{\tau_n}$ are the uncertain and nominal motor torque coefficients, respectively; $q$, $\dot{q}$, $\ddot{q}$ are the angle, velocity, and acceleration of motor shaft, respectively; $\omega_s = T_s^{-1}$ is the sampling frequency; $g_{DOb}$ is the bandwidth of the low-pass-filter; $I$ is the motor thrust current; $\eta$ is noise; $ZoH$ is Zero-order Hold; $\tau_d$ is the external disturbance torque; $\tau_{dis}$ is the internal and external disturbance torque; $\hat{\bullet}$ is the estimation of $\bullet$; and $\bullet_{des}$ is the desired $\bullet$.

To estimate the internal and external disturbances of a motion control system, a DOb is synthesised using the identified (i.e., nominal) system model, a low-pass-filter, and the velocity measurement of a servo system as illustrated in Fig. 1. The transfer functions between the exogenous inputs and output are derived from this figure as follows.

Continuous-time Domain:

$$\dot{q}(s) = \alpha \frac{s+g_{DOb}}{s+\alpha g_{DOb}} \dot{q}_{des}(s) - \frac{1}{J_m s} \mathrm{S}_{\mathrm{DOb}}(s) \tau_d(s) - \mathrm{T}_{\mathrm{DOb}}(s) \eta(s) \quad (1)$$

where $\mathrm{S}_{\mathrm{DOb}}(s) = \frac{1}{1+\mathrm{L}_{\mathrm{DOb}}(s)} = \frac{s}{s+\alpha g_{DOb}}$ and $\mathrm{T}_{\mathrm{DOb}}(s) = 1 - \mathrm{S}_{\mathrm{DOb}}(s)$ $\frac{\mathrm{L}_{\mathrm{DOb}}(s)}{1+\mathrm{L}_{\mathrm{DOb}}(s)} = \frac{\alpha g_{DOb}}{s+\alpha g_{DOb}}$ are the sensitivity and complementary sensitivity transfer functions in which $\mathrm{L}_{\mathrm{DOb}}(s) = \alpha g_{DOb}/s$ is the open-loop transfer function of the inner-loop and $\alpha = \left(J_{m_n} K_\tau\right) / \left(J_m K_{\tau_n}\right)$.

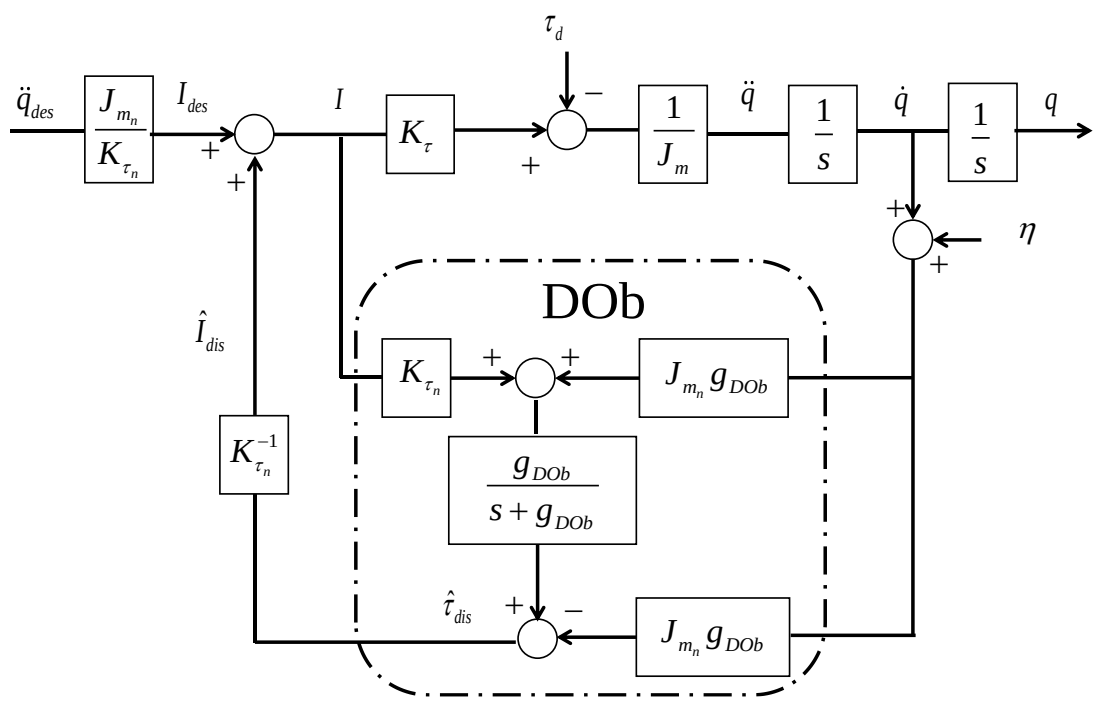

a) Continuous-time Domain.

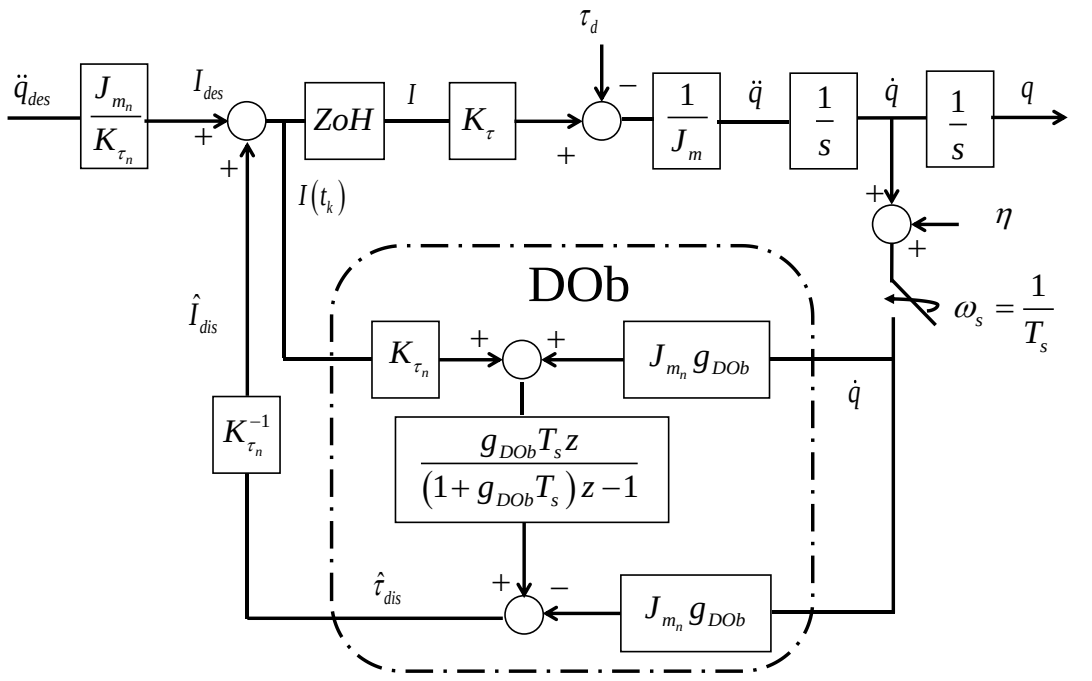

b) Discrete-time Domain.

Fig. 1: Block diagrams of the Disturbance Observer.

Discrete-time Domain:

$$\dot{q}(z) = \alpha \frac{(1+g_{DOb}T_s)z-1}{z-(1-\alpha g_{DOb}T_s)} \frac{T_s}{z-1} \ddot{q}_{des}(z) - \frac{1}{J_m} \mathrm{S}_{\mathrm{DOb}}(z) \mathrm{G}_{\mathrm{V}}(z) \tau_d(z) - \mathrm{T}_{\mathrm{DOb}}(z)\eta(z) \quad (2)$$

where $\mathrm{S}_{\mathrm{DOb}}(z) = \frac{1}{1+\mathrm{L}_{\mathrm{DOb}}(z)} = \frac{z-1}{z-(1-\alpha g_{DOb}T_s)}$ and $\mathrm{T}_{\mathrm{DOb}}(z) = 1 - \mathrm{S}_{\mathrm{DOb}}(z)$ $= \frac{\mathrm{L}_{\mathrm{DOb}}(z)}{1+\mathrm{L}_{\mathrm{DOb}}(z)} = \frac{\alpha g_{DOb}T_s}{z-(1-\alpha g_{DOb}T_s)}$ are the sensitivity and complementary sensitivity transfer functions in which $\mathrm{L}_{\mathrm{DOb}}(z) = \frac{\alpha g_{DOb}T_s}{z-1}$ is the open-loop transfer function of the inner-loop.

Equations (1) and (2) show that the robust stability and performance of the DOb can be adjusted by tuning the nominal plant parameters and the bandwidth of the DOb. For example, using the higher/lower values of inertia/torque coefficient can improve the stability by increasing the phase-lead effect in the inner-loop. Also, increasing the bandwidth of the DOb improves the robustness against disturbances as $\mathrm{S}_{\mathrm{DOb}}(\bullet)$ gets smaller at low frequency range. This, however, deteriorates the noise-sensitivity as $\mathrm{T}_{\mathrm{DOb}}(\bullet)$ gets larger at high frequency range. The sensitivity and complementary sensitivity functions are very useful analysis and synthesis tools which we use in deriving the fundamental design constraints of the DOb in Section III.

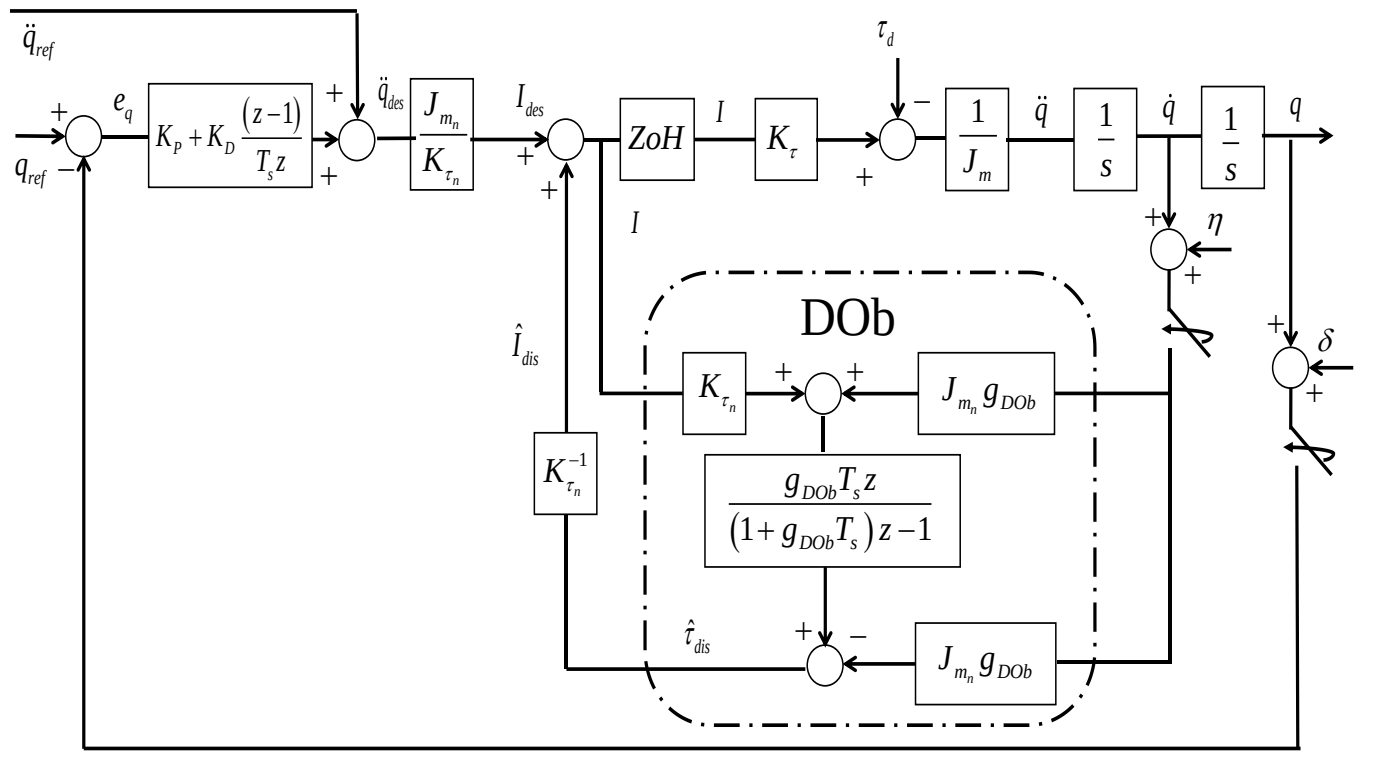

Fig. 2: Block diagram of the DOb-based Robust Position Controller.

In the last two decades, various applications of this disturbance estimation tool, spaning from precise positioning of industrial machines to estimating time-delay in communication, have been proposed in the literature [4, 9]. The following subsections introduce the most frequently used DOb-based robust motion control systems, viz, the DOb-based robust position controller and the DOb-based robust force controller.

### *B. Disturbance Observer-based Robust Position Control*

The block diagram of the DOb-based robust position controller is illustrated in Fig. 2. In this figure, $K_P$ and $K_D$ represent the proportional and derivative gains of the outer-loop PD controller, respectively, $\delta$ represents the noise of position measurement, and $\bullet_{ref}$ represents the reference signal of $\bullet$. The other parameters are same as defined earlier. For the sake of brevity, we only introduce the robust position controller in the discrete-time domain. Readers are referred to [29] for the continuous-time analysis and synthesis of the DOb-based robust position controller.

The robust position controller consists of two feedback loops: an inner-loop in which the robustness of the controller is adjusted by the DOb and an outer-loop in which the performance of the controller is adjusted by the proportional and derivative control gains. The robustness of the position control system is intuitively achieved by feeding back the estimation of disturbances in the inner-loop. The outer-loop performance controller can be freely tuned by considering only the nominal plant parameters because disturbances, such as load, friction and parameter variations, are precisely suppressed in the inner-loop. Since the robustness and performance can be independently adjusted in the inner- and outer- loops, the DOb-based robust position controller has two-degrees-of-freedom.

### *C. Disturbance Observer-based Robust Force Control*

Figure 3 illustrates the block diagram of the DOb-based robust force controller. In this figure, $K_{env}$ and $D_{env}$ represent the stiffness and damping of environment, respectively, $g_{RTOb}$ represents the bandwidth of the RTOb, $\bullet_i$ represents the identified $\bullet$, and $C_\tau$ represents the torque control gain. The other parameters are same as defined earlier.

Similar to the robust position controller, the DOb-based robust force controller is synthesised by employing two feedback control loops. A DOb is employed in the inner-loop to improve the robustness of the force controller by suppressing disturbances. Another DOb (viz. RTOb) is employed in the outer-loop to estimate interaction torques and improve the performance of the torque control system.

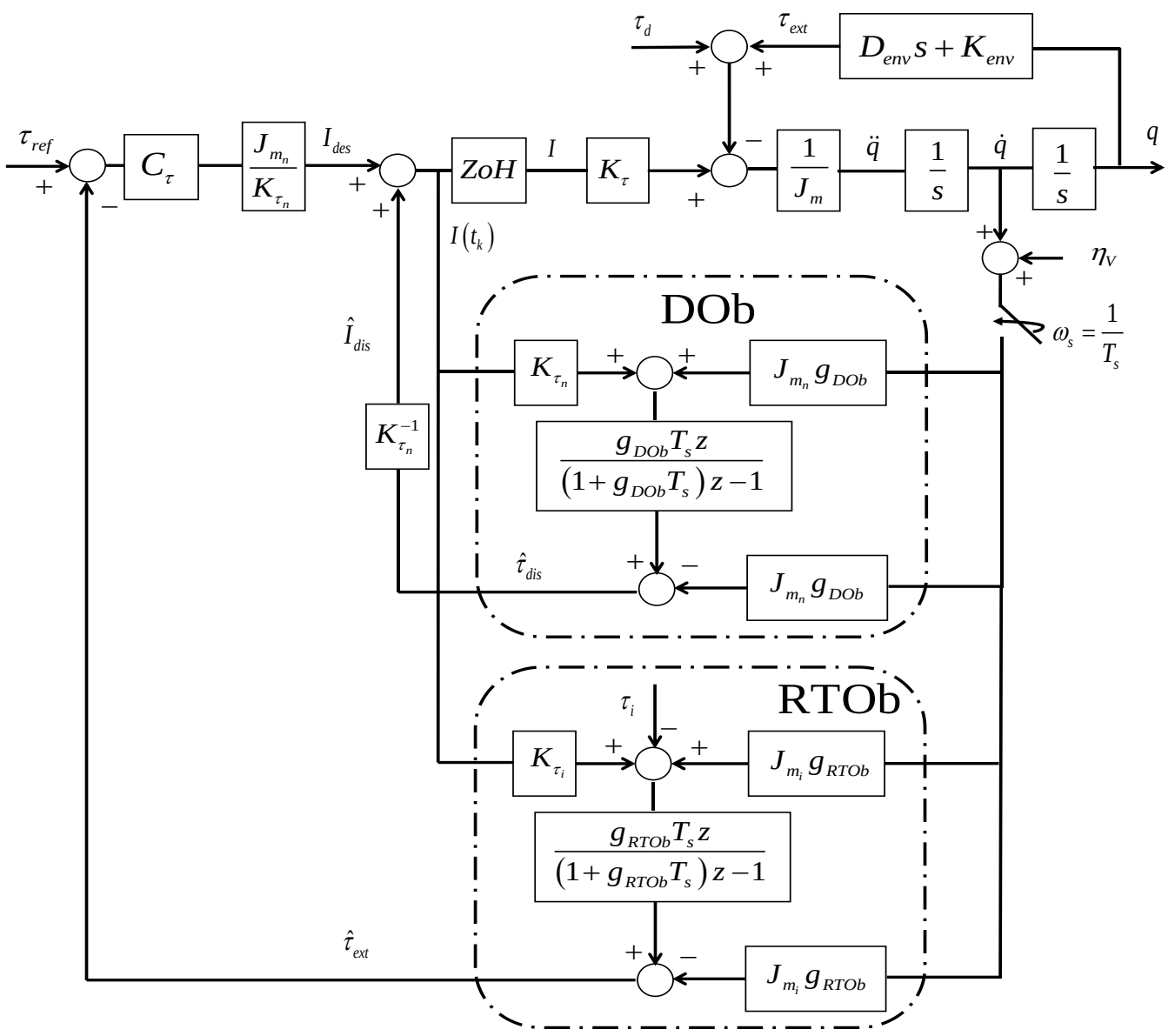

Fig. 3: Block diagram of the DOb-based Robust Force Controller.

The superiorities of the DOb-based robust force controller over other force control methods have been reported in the literature. For example, the robust force controller can precisely track force trajectories using the DOb in the inner-loop, enabling real-haptic sensation in teleopearation. Also, the DOb-based robust force controller notably improves the stability of contact motion by estimating interaction forces within a large frequency range via RTOb in the outer-loop. Reader is referred to [8, 26] for further details on the DOb-based robust force control systems.

## III. Design Constraints of the Disturbance Observer-based Robust Motion Control Systems

Frequency responses of the sensitivity and complementary sensitivity functions have been widely used in adjusting the control parameters of the DOb-based robust motion control systems. For example, a well-known design trade-off between the robustness against disturbances and the noise-sensitivity of the DOb can be easily shown using the frequency responses of the sensitivity and complementary sensitivity functions given in Eq. (1). The bandwidth of the DOb is generally set as high as possible by considering the noise-sensitivity of the velocity measurement system in practice.

The frequency domain analysis and synthesis of the DOb-based robust digital motion control systems are generally conducted in the continuous-time domain due to simplicity. This section shows that this simplification may cause severe robust stability and performance problems in practice by employing the Bode Integral Theorem in the continuous- and discrete- time domains.

### *A. Continuous-Time Analysis of the DOb*

Let us start with deriving the fundamental design constraints of the DOb-based robust motion control systems using the Bode Integral Theorem in the continuous-time domain.

When the generalised Bode Integral Theorem is applied to the DOb-based robust motion controller illustrated in Fig. 1a, the following equation is obtained.

$$\int_0^\infty \log\left(\left|\mathrm{S}_{\mathrm{DOb}}\left(j\omega\right)\right|\right)d\omega = \pi\sum_k \mathrm{Re}\left(p_{u_k}\right) - \frac{\pi}{2}\lim_{s\to\infty} s\mathrm{L}_{\mathrm{DOb}}\left(s\right) = -\frac{\pi}{2}\alpha g_{DOb} \quad (3)$$

where $\omega$ is frequency, $j^2 = -1$ is a complex number, $p_{u_k}$ is the k[th] unstable pole of the open-loop transfer function [28]. The continuous open-loop and sensitivity transfer functions, i.e., $\mathrm{L}_{\mathrm{DOb}}(s)$ and $\mathrm{S}_{\mathrm{DOb}}(s)$, are given in Eq. (1).

Equations (1) and (3) show that as the bandwidth of the DOb and/or the control parameter $\alpha$ increases, the right hand side of the Bode's integral equation gets smaller. Therefore, without suffering from a high-sensitivity peak, one can freely improve the robustness against disturbances and the phase-lead effect of the inner-loop by increasing $g_{DOb}$ and $\alpha$, respectively. In other words, Eq. (3) shows that the DOb-based robust motion controller is not subject to the waterbed effect and one can achieve good robust stability and performance for all values of the DOb's design parameters.

This, however, is not what we observe when conducting robust motion control experiments in practice. It is a well-known fact that not only the noise sensitivity but also the stability of the robust motion controller deteriorates as the bandwidth and nominal inertia of the DOb are increased [21 – 23]. To conduct high-performance robust motion control applications in practice, it is essential to derive the design constraints on the bandwidth and nominal plant parameters of the DOb.

*B. Discrete-Time Analysis of the DOb*

Let us now derive the fundamental design constraints of the DOb-based robust motion control systems in the discrete-time domain.

The Bode's integral equation of the DOb-based robust motion controller illustrated in Fig. 1b is as follows.

$$\int_{-\pi}^{\pi} \ln\left|\mathrm{S}_{\mathrm{DOb}}\left(\mathrm{e}^{j\omega T_s}\right)\right| d\omega T_s = 2\pi\left(\ln\sum\left|p_{u_k}\right| - \ln\left|1+\lim_{z\to\infty}\mathrm{L}_{\mathrm{DOb}}\left(z\right)\right|\right) = 0 \quad (4)$$

where $z = \mathrm{e}^{j\omega T_s}$ is a complex number [28]. The discrete open-loop and sensitivity transfer functions, i.e., $\mathrm{L}_{\mathrm{DOb}}(z)$ and $\mathrm{S}_{\mathrm{DOb}}(z)$, are given in Eq. (2).

Compared to Eq. (3), the interval of the Bode's sensitivity integral is finite, [-π, π], in Eq. (4). This causes a stricter design constraint in digital feedback control systems. Moreover, the right hand side of Eq. (4) is zero. Therefore, to hold the Bode's integral equation, the peak of the frequency response of the sensitivity function $\left|\mathrm{S}_{\mathrm{DOb}}\left(\mathrm{e}^{j\omega T_s}\right)\right|$ should increase at medium/high frequencies as the sensitivity function decreases at low frequency range by using higher values of $g_{DOb}$ and/or $\alpha$. In other words, the digital DOb-based robust motion controller is subject to the waterbed effect; therefore, good robust stability and performance cannot be achieved for all values of $\alpha$ and $g_{DOb}$ in practical motion control applications.

This simple analysis clearly shows that the design constraints on the nominal plant model and the bandwidth of the DOb become stricter when implementing digital controllers. The analysis and synthesis conducted in the continuous-time domain may cause severe stability and performance problems in practice. Therefore, we need to employ discrete-time analysis and synthesis methods in implementing the DOb-based robust motion control systems.

The proposed qualitative analysis, however, is not sufficient in analytically deriving the design constraints of the DOb-based robust motion control systems. Therefore, we need to put more effort to find the upper and lower bounds on the nominal plant model and the bandwidth of the DOb. Reader is referred to [27] for an example of analytical design constraints of the DOb in the discrete-time domain.

*C. Design Constraints of the Robust Motion Controller*

To conduct high-performance robust motion control applications, not only the inner-loop but also the outer-loop should be properly tuned. The Bode's sensitivity integral can be similarly used to derive the fundamental design constraints of the DOb-based robust motion control systems in the outer-loop.

Let us consider the robust position controller illustrated in Fig. 2. The transfer functions between the exogenous inputs and output can be directly derived from this figure as follows:

$$\begin{aligned} q(z) &= \mathrm{L}_{\mathrm{PC}}(z)e_q(z) + \frac{1}{J_m}\mathrm{S}_{\mathrm{DOb}}(z)\mathrm{S}_{\mathrm{PC}}(z)\tau_d(z) + \\ &\quad \mathrm{T}_{\mathrm{DOb}}(z)\mathrm{T}_{\mathrm{PC}}(z)\frac{T_s}{2}\frac{z+1}{z-1}\eta(z) + \mathrm{T}_{\mathrm{PC}}(z)\delta(z) \end{aligned} \quad (5)$$

where $\mathrm{S}_{\mathrm{PC}}(z) = \frac{1}{1+\mathrm{L}_{\mathrm{PC}}(z)}$ and $\mathrm{T}_{\mathrm{PC}}(z) = 1 - \mathrm{S}_{\mathrm{PC}}(z) = \frac{\mathrm{L}_{\mathrm{PC}}(z)}{1+\mathrm{L}_{\mathrm{PC}}(z)}$ are the sensitivity and complementary sensitivity transfer functions of the outer-loop position controller in which the open-loop transfer function $\mathrm{L}_{\mathrm{PC}}(z)$ is given in Eq. (6).

$$\mathrm{L}_{\mathrm{PC}}(z) = \left(K_P + K_D\frac{z-1}{T_s z}\right)\left(\alpha\frac{(1+g_{DOb}T_s)z-1}{z-(1-\alpha g_{DOb}T_s)}\right)\frac{T_s^2}{2}\frac{z+1}{(z-1)^2} \quad (6)$$

The generalised Bode's integral equation of the DOb-based robust position controller illustrated in Fig. 2 is as follows.

$$\int_{-\pi}^{\pi} \ln\left|\mathrm{S}_{\mathrm{PC}}\left(\mathrm{e}^{j\omega T_s}\right)\right| d\omega T_s = 2\pi\left(\ln\sum\left|\lambda_{u_k}\right| - \ln\left|1+\lim_{z\to\infty}\mathrm{L}_{\mathrm{PC}}(z)\right|\right) = 2\pi\ln\left|\lambda\right| \quad (7)$$

where $\lambda_{u_k}$ is the k[th] unstable pole of the open-loop transfer function $\mathrm{L}_{\mathrm{PC}}(z)$ [28].

Equation (7) shows that the DOb-based robust position controller is subject to the waterbed effect in the outer-loop. This design constraint becomes stricter when the inner-loop control parameters are not properly tuned so that $\mathrm{L}_{\mathrm{PC}}(z)$ has an unstable pole. Although this analysis provides a clear insight into the fundamental design constraints of the DOb-based robust position controller, it is still a very challenging task to obtain the upper and lower limits of the control parameters using Eq. (7). Therefore, more effort should be made to establish analytical design tools for the DOb-based robust motion control systems.

For the sake of brevity, the robust force controller is not considered in this section. However, the design constraints of the DOb-based robust force controller can be similarly analysed using the Bode's Integral Theorem [30].

## IV. Simulations and Experiments

This section verifies the proposed analysis method by providing simulation and experimental results.

Let us first analyse the robust stability and performance of the DOb. Figure 4 illustrates the frequency responses of the inner-loop's sensitivity and complementary sensitivity functions in the continuous- and discrete- time domains. As shown in Figure 4a, continuous-time analysis shows that good robust stability and performance can be achieved for all values of $\alpha$ and $g_{DOb}$. Therefore, we can adjust the robustness and phase-lead of the inner-loop by only considering the noise sensitivity, i.e., high-frequency response of the complementary sensitivity function. This widely adopted analysis, however, cannot reflect the dynamic behaviour of the robust motion control systems in practice. In fact, the robust stability and performance of the DOb deteriorate due to the discrete-time design constraints as shown in Fig. 4b. To achieve good robust stability and performance, we

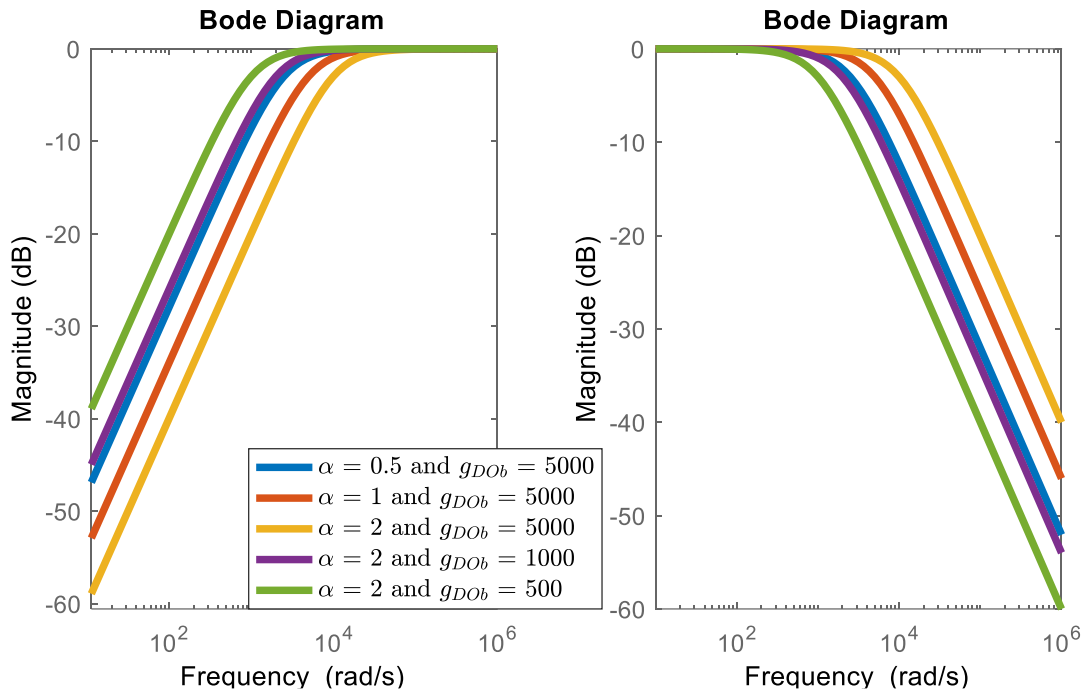

a) Frequency responses in the continuous-time domain.

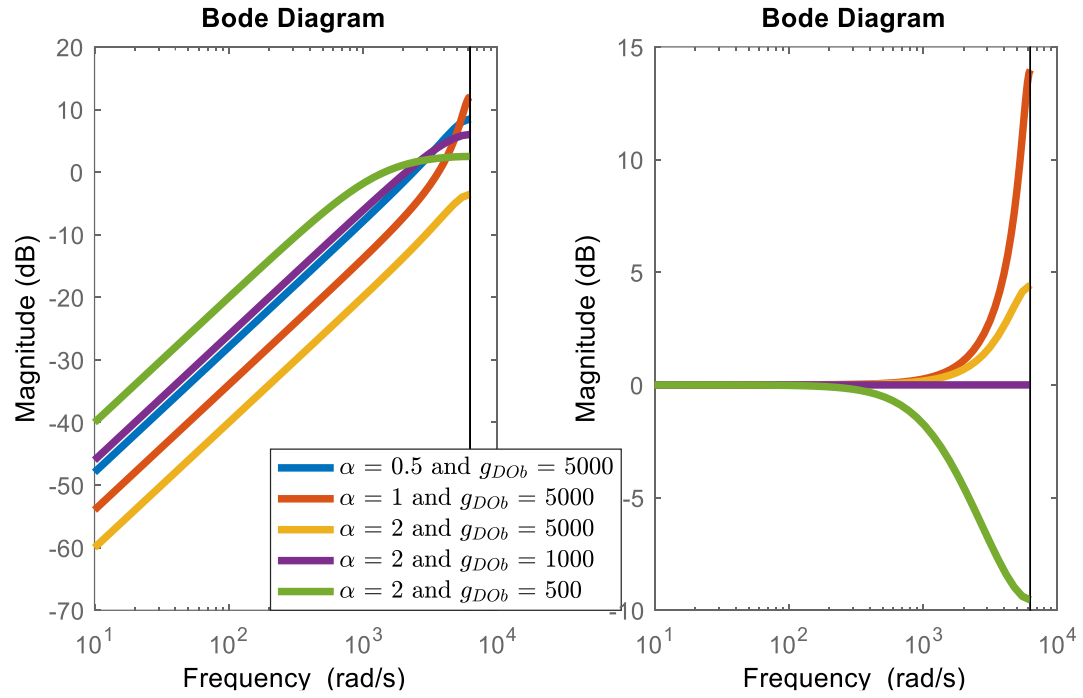

b) Frequency responses in the discrete-time domain when $T_s$ = 0.5 ms

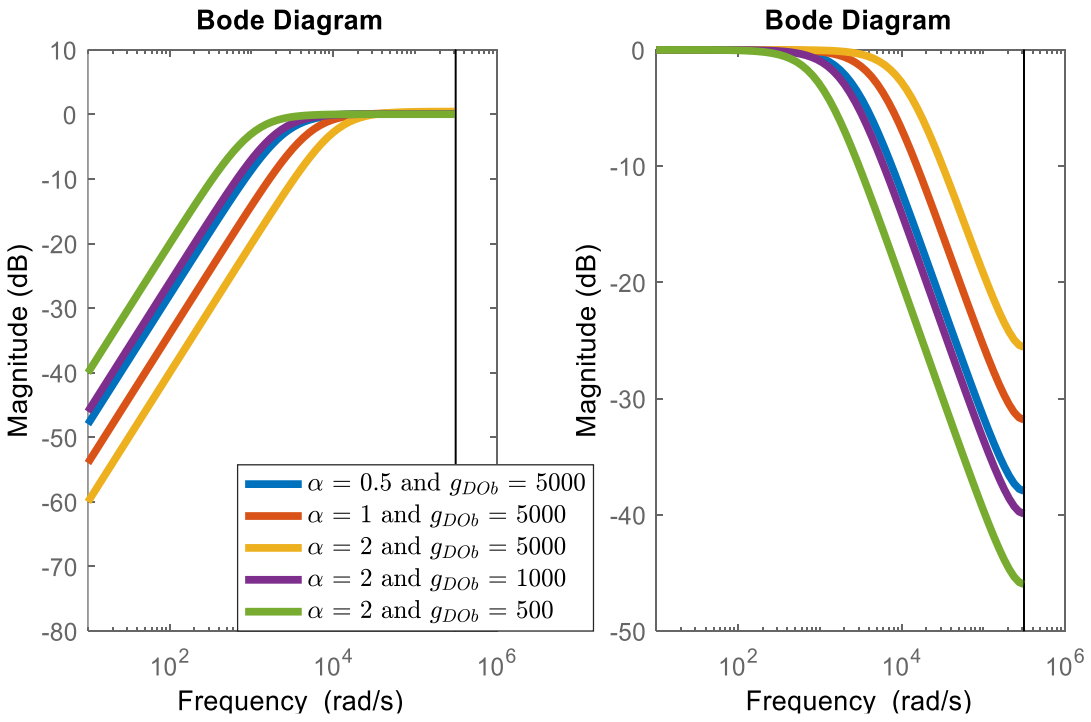

b) Frequency responses in the discrete-time domain when $T_s$ = 100 μs

Fig. 4. Inner-loop sensitivity (left-figures) and complementary sensitivity (right-figures) functions' frequency responses.

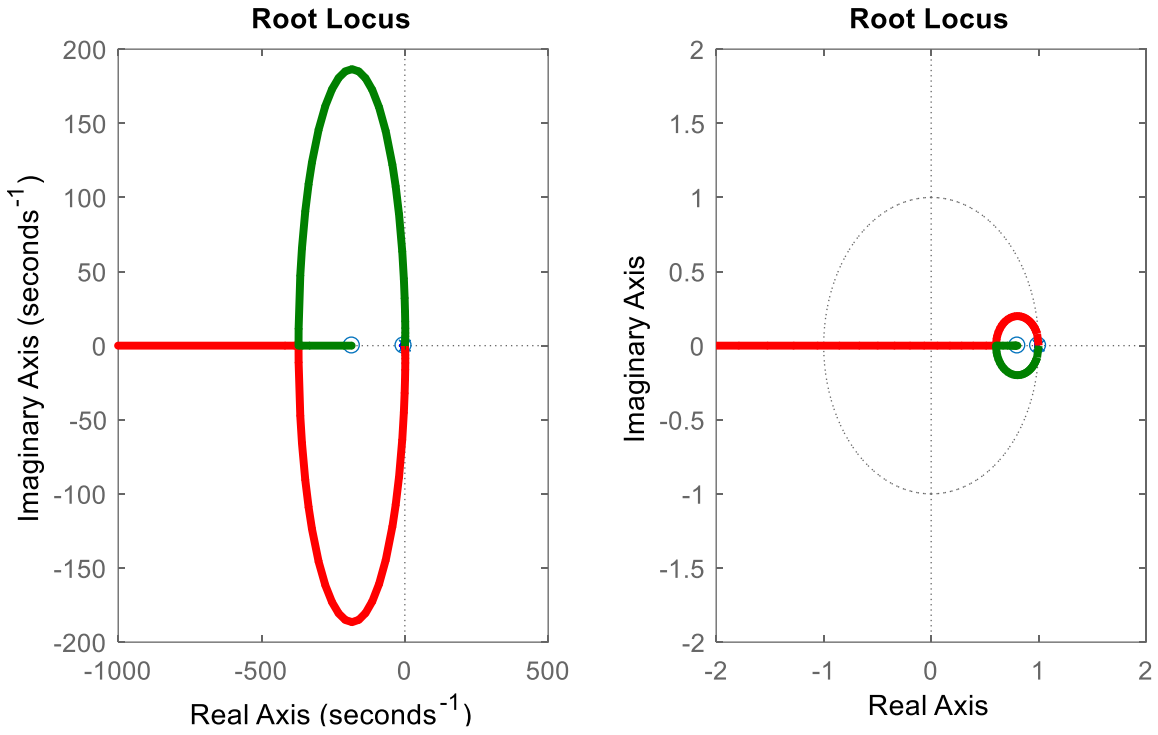

a) Root loci of the robust position controller with respect to $g_{DOb}$.

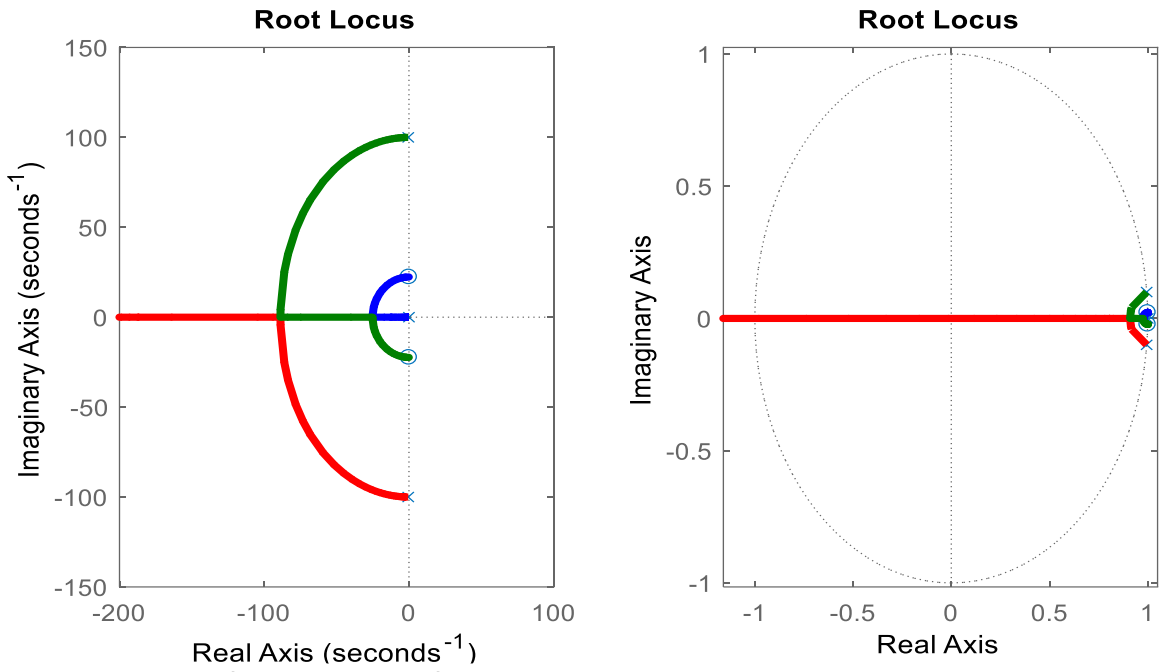

b) Root loci of the robust force controller with respect to $g_{DOb}$.

Fig. 5. Root-loci of the robust motion control systems in the continuous-time domain (left figure) and discrete-time domain (right figure).

need to limit either the robustness against disturbances or the phase-lead effect in the inner-loop using low values of $g_{DOb}$ and $\alpha$ in the design of the DOb, respectively. Figure 4c shows that the design constraints on the bandwidth of the DOb and nominal plant model can also be relaxed by decreasing the sampling time. This, however, generally increases cost.

Figure 5 illustrates the root-loci of the robust motion control systems with respect to $g_{DOb}$ in the continuous- and discrete-time domains. It is assumed $J_m$ = 0.01, and the environmental stiffness and damping are $K_{env}$ = 10000 and $D_{env}$ =1, respectively. The robust position controller is synthesised using, $K_P$ = 5000 and $K_D$ = 25, and the robust force controller is tuned using the exact inertia and torque coefficient values in the DOb and RTOb synthesis, $C_f$ = 0.5, and $g_{DOb}$ = $g_{RTOb}$. Right figures in Fig. 5 clearly show that the stability of the DOb-based digital robust motion control systems deteriorates as the bandwidth of the DOb increases. This, however, cannot be observed when the stability analysis is conducted in the continuous-time domain (see left figures in Fig. 5).

It is noted that the stability of the robust motion control systems can be improved by properly tuning the design parameters of the DOb and RTOb. For example, the stability of the robust position controller is improved using higher values of the nominal inertia, i.e., α, in the design of the DOb [29], and the stability of the robust force controller is improved using lower values of the identified inertia in the design of the RTOb [8]. However, the asymptotic behaviours of the robust motion controllers are similar to Fig. 5.

Last, let us verify the proposed analysis method with an experiment. Figure 6 illustrates the robust position control experiment when different values of $g_{DOb}$ are used in the DOb synthesis. It is clear from this figure that the stability of the robust position controller deteriorates when the bandwidth of the DOb does not satisfy the robust stability constraint described in Section III.

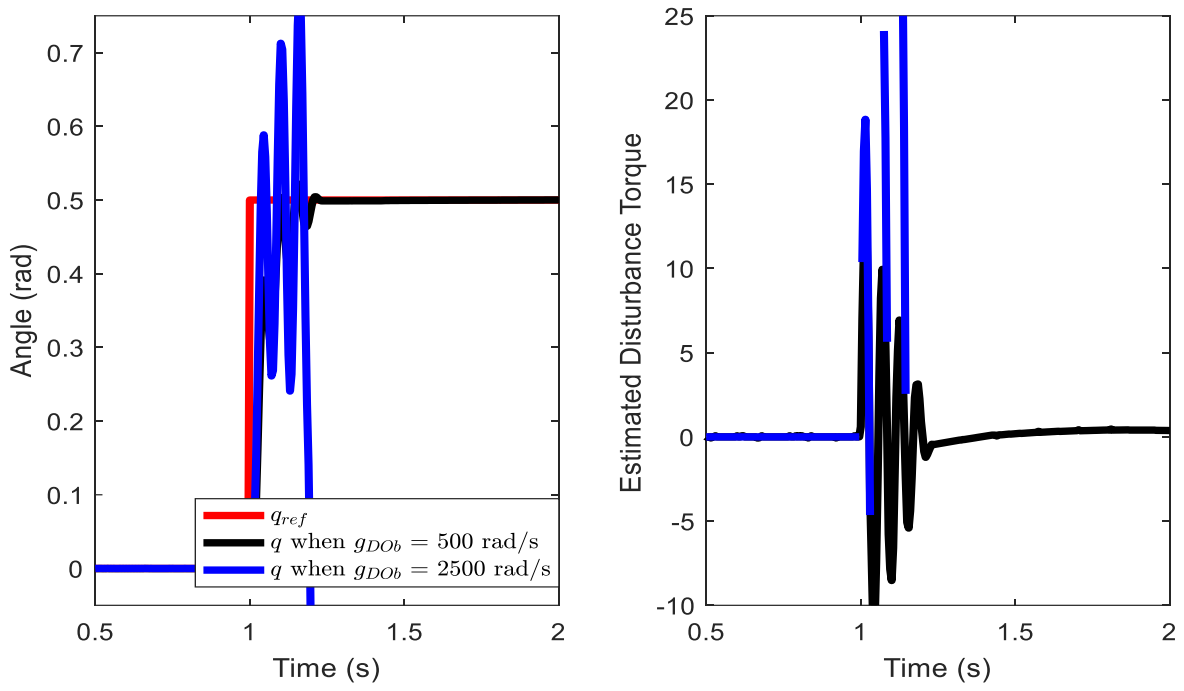

Fig. 6. Robust position control experiment.

## V. Conclusions

This paper shows that the design constraints on the nominal plant model and the bandwidth of the DOb become stricter when the robust motion controllers are synthesised using digital systems such as computers and microcontrollers. This limits the robustness against disturbances and flexibility in adjusting the nominal plant model in the DOb-based control. When the design constraints are violated, not only the noise sensitivity but also the robust stability and performance may notably deteriorate. This, however, cannot be explained in the continuous-time domain. Therefore, this paper recommends discrete-time analysis and synthesis methods for the DOb-based robust motion control systems.